\documentclass[epj,nopacs]{svjour}
\usepackage{graphicx,amssymb}
\usepackage[intlimits]{amsmath}

\begin{document}

\title{Plasmaneutrino spectrum}

\author{A. Odrzywo\l{}ek}

\institute{M. Smoluchowski Institute of Physics, 
Jagiellonian University, Reymonta 4, 30-059 Krakow, Poland}

\date{\today}
\abstract{
Spectrum of the neutrinos produced in the massive photon
and longitudal plasmon decay process has been computed
with four levels of approximation for the dispersion
relations. Some analytical formulae in limiting cases
are derived. Interesting conclusions related to previous
calculations of the energy loss in stars are presented.
High energy tail of the neutrino spectrum is shown
to be proportional to exp(-E/kT), where E is the neutrino
energy and kT is the temperature of the plasma.
}

\maketitle

\begin{section}{Introduction \& Motivation}

Thermal neutrino loses from plasma are
very important for stellar astrophysics \cite{Arnett,Bisnovaty}.
Plasmon decay is one of the three main reactions.
Extensive calculations for these processes were
done by group of Itoh 
\cite{Itoh_I,Itoh_I_erratum,Itoh_II,Itoh_III,Itoh_III_erratum,Itoh_IV,Itoh_V,Itoh_VI,Itoh_VII}.  
Other influential article include 
\cite{BPS,Adams-Woo,Dicus,BraatenSegel,BraatenPRL,Schinder,BlinnikovRudzskij,BlinnikovRudzskij2,Raffelt}. Meanwhile, our abilities
to detect neutrinos has grown by many orders of magnitude,
beginning with $1.4$ tonne experiment of Reines\&Cowan \cite{ReinesCowan} up to
the biggest existing now $50$~kt Super-Kamiokande detector \cite{SK}. 
Recently, ''GADZOOKS!'' upgrade to Super-Kamiokande proposed by Beacom\&Vagins \cite{Gadzooks} 
attract attention of both experimental and theoretical physicists. At last
one new source of the astrophysical antineutrinos is guaranteed
with this upgrade, namely Diffuse Supernova Neutrino Background \cite{SN1987A-20th,DSNB}.
Pre-supernova stars will be available to observations
out to $\sim$2 kiloparsecs \cite{SN1987A-20th}.
This technique is the only extensible to megaton scale \cite{SN1987A-20th}.
Memphys, Hyper-Kamiokande and UNO (Mt-scale water Cherenkov detectors cf. e.g.
\cite{Fogli})
proposals now seriously consider to add GdCl$_3$ to the one of the tanks with 
typical three-tank design \cite{NNN06}.
Recently, the discussion on the geoneutrino detection \cite{Learned_GEO}, 
increased attention
to the deep underwater neutrino observatories \cite{HanoHano} with target mass 
5-10 Mt \cite{SN1987A-20th} and even bigger \cite{GigatonArray}.
It seems that (anti)neutrino astronomy is on our doorstep,
but numerous astrophysical sources of the $\nu$'s still are
not analyzed from the detection point of view.

Detection of the solar \cite{Davis,Gallex,SNO,SK_sun} and supernova neutrinos 
\cite{SK_sn,IMB,LSD,Baksan} 
was accompanied and followed with extensive set of detailed 
calculations (see e.g. \cite{Bahcall,MPA,Burrows,Mezzacappa,Yamada,Bethe}
and references therein as a representatives of this broad subject)
of the neutrino spectrum. On the contrary, very little
is known about spectral neutrino emission from other astrophysical
objects. Usually, some analytical representation of the spectrum
is used, based on earlier experience 
and numerical simulations, cf. e.g. \cite{Pons}.
While this approach is justified for supernovae, where neutrinos
are trapped, other astrophysical
objects are transparent to neutrinos, and spectrum can be
computed with an arbitrary precision. 
Our goal is to compute neutrino spectra as exact as possible and fill this gap. 
Plasmaneutrino process dominates dense, degenerate objects
like red giant cores \cite{RedGiants}, cooling white dwarfs \cite{WDcool} including 
Ia supernova progenitors before so-called ,,smoldering'' phase \cite{IaSmouldering}. It is also 
important secondary cooling process in  e.g. neutron star crusts \cite{HaenselRev}
and massive stars \cite{Heger_rev}.
Unfortunately, thermal neutrino loses
usually are calculated using methods completely erasing almost any
information related to the neutrino energy $\mathcal{E}_\nu$ and directionality
as well. This information is not required to compute total energy $Q$ radiated
as neutrinos per unit volume and time.
From experimental point of view, however, it is extremely
important if given amount of energy is radiated as e.g. numerous
keV neutrinos or one 10 MeV neutrino. In the first case
we are unable to detect (using available techniques) any transient neutrino source 
regardless of the total luminosity and proximity of the object. In the second
case we can detect astrophysical neutrino sources if they are
strong and not too far away using advanced detector which is big enough.  
 
Few of the research articles in this area attempt to estimate
average neutrino energy \cite{BraatenPRL,Schinder,Ratkovic,Dutta} 
computing additionally reaction rate $R$. Strangely, they presented
figures and formulae for $Q/R$ instead of $\frac{1}{2} Q/R$. This gives
false picture of real situation, as former expression gives 
$\langle \mathcal{E}_{\nu}+\mathcal{E}_{\bar{\nu}} \rangle$. 
Obviously, we detect {\em neutrinos} not $\nu$-$\bar{\nu}$ pairs.
$\frac{1}{2} Q/R$ {\em do not} 
give average neutrino energy, as in general neutrino
and antineutrino spectra are different. As we will see 
{\em only} for longitudal plasmon decay neutrinos energies
of neutrinos and antineutrinos are equal. However, difference
in all situations where thermal neutrino loses are important is numerically
small and formula: 
\begin{equation}
\langle \mathcal{E}_{\nu} \rangle \simeq \frac{1}{2} \frac{Q}{R}
\end{equation}
is still a ''working'' estimate.

Mean neutrino energy is useful in the purpose of qualitative discussion
of the detection prospects/methods. Quantitative discussion require knowledge
of spectrum shape (differential emissivity $dR/d\mathcal{E}_\nu$). 
High energy tail is particularly important from an experimental
detection point of view.  Detection of the lowest energy neutrinos is extremely
challenging due to numerous background signal noise sources e.g. $^{14}$C
decay for $\mathcal{E}_\nu<200$~keV \cite{14C}.
Relevant calculations for the spectrum  of the medium energy
$\langle \mathcal{E}_{\nu} \rangle \sim 1$MeV
neutrinos emitted from thermal processes has become available
recently \cite{Ratkovic,Dutta,MOK}. Purpose of this article is
to develop accurate methods and discuss various theoretical and practical 
(important for detection) aspects of the neutrino spectra from astrophysical plasma process.
This could help experimental physicists to discuss possible realistic
approach to detect astrophysical sources of the neutrinos in the future.

\end{section}

\begin{section}{Plasmaneutrino spectrum}

\begin{subsection}{Properties of plasmons}

Emissivity and the spectrum shape from the plasmon decay is strongly affected by the dispersion relation
for transverse plasmons (massive in-medium photons) and longitudal
plasmons. In contrast to transverse plasmons, with vacuum dispersion relation
$\omega(k)=k$, longitudal plasmons exist only in the plasma. 
Dispersion relation, by the definition is a function
 $\omega(k)$ where $\hbar \omega$ is the energy of the (quasi)particle
and $\hbar k$ is the momentum.
Issues related to particular handling of these functions
are discussed clearly in the article of Braaten and Segel \cite{BraatenSegel}. 
We will repeat here the most important features of the plasmons.

For both types, plasmon energy for momentum $k=0$ is equal to $\omega_0$. 
Value $\omega_0 \equiv \omega(0)$ is refereed to as {\em  plasma frequency} 
and can be computed from:
\begin{equation}
  \label{plasma_frequency}
  \omega_0^2 = \frac{4 \alpha}{\pi} \int_0^{\infty} \frac{p^2}{E}
    \left( 1 - \frac{v^2}{3} \right) (f_1 + f_2) \; dp
\end{equation}
where $v=p/E$, $E = \sqrt{p^2 + m_e^2}$ ($\hbar=c=1$ units are used),
$m_e\simeq0.511$~MeV and
fine structure constant is $\alpha=1/137.036$ \cite{PDBook}.
Functions $f_1$ and $f_2$
are the Fermi-Dirac distributions for electrons and positrons, respectively:
\begin{equation}
\label{F-D}
f_1 = \frac{1}{e^{(E - \mu)/kT}+1}, \qquad
f_2 = \frac{1}{e^{(E + \mu)/kT}+1}.
\end{equation}
Quantity $\mu$ is the electron chemical potential (including the rest mass).
Other important parameters include first relativistic correction
$\omega_1$:
\begin{equation}
\omega_1^2 = \frac{4 \alpha}{\pi} \int_0^{\infty} \frac{p^2}{E}
\left( \frac{5}{3} v^2 - v^4 \right) (f_1 + f_2) \; dp
\end{equation} 
maximum longitudal plasmon momentum (energy) $k_{max}$:
\begin{equation}
\label{max_momentum}
k_{max}^2 \equiv \omega_{max}^2  = \frac{4 \alpha}{\pi} \int_0^{\infty} \frac{p^2}{E}
\left( \frac{1}{v} \ln{\frac{1-v}{1+v}} - 1 \right) (f_1 + f_2) \; dp
\end{equation} 
and asymptotic transverse plasmon mass $m_t$:
\begin{equation}
\label{tmass}
m_t^2 = \frac{4 \alpha}{\pi} \int_0^{\infty} \frac{p^2}{E}
 (f_1 + f_2) \; dp.
\end{equation} 
Value $m_t$ is often referred to as thermal photon mass. 
We also define parameter $v_\ast$:
\begin{equation}
\label{electron_velocity}
v_\ast = \frac{\omega_1}{\omega_0}
\end{equation}
interpreted as typical velocity of the electrons in the plasma \cite{BraatenSegel}.
Axial polarization coefficient is:
\begin{equation}
\omega_A = \frac{2 \alpha}{\pi} \int_0^\infty 
 \frac{p^2}{E^2}
 \left( 1 - \frac{2}{3} \, v^2 \right)
 (f_1 - f_2)
\; dp.
\end{equation}
Value of the $\omega_A$ is a measure of the difference between
neutrino and antineutrino spectra. 
Set of numerical values used
to display sample result
is presented in Table~\ref{tbl1}.

\begin{table}
\caption{\label{tbl1} Plasma properties for typical
massive star during Si burning. All values in MeV.
}
\begin{tabular*}{\columnwidth}{cc|ccccc}
$kT$ & $\mu$ & $\omega_0$ & $\omega_1$ & $m_t$ & $\omega_{max}$ & $\omega_A$\\
 \hline
0.32 & 1.33 & 0.074 & 0.070 & 0.086 & 0.133 & 0.002
\end{tabular*}
\end{table}

Values $\omega_0, \omega_{max}, m_t$ define sub-area of the $\omega$-$k$
plane where dispersion relations for photons $\omega_t(k)$
and longitudal plasmons $\omega_l(k)$ are found:
\begin{subequations}
\label{sub-area}
\begin{equation}
\max{(k,\omega_0)} \leq \omega_l(k) \leq \omega_{max}, \quad 0 \leq k \leq k_{max}
\end{equation}
\begin{equation}
\sqrt{k^2+\omega_0^2} \leq \omega_t(k) \leq \sqrt{k^2+m_t^2}, \quad  \quad 0 \leq k \leq \infty
\end{equation}
\end{subequations} 

\begin{figure*}
\begin{tabular}{cc}
  \includegraphics[width=0.5\textwidth]{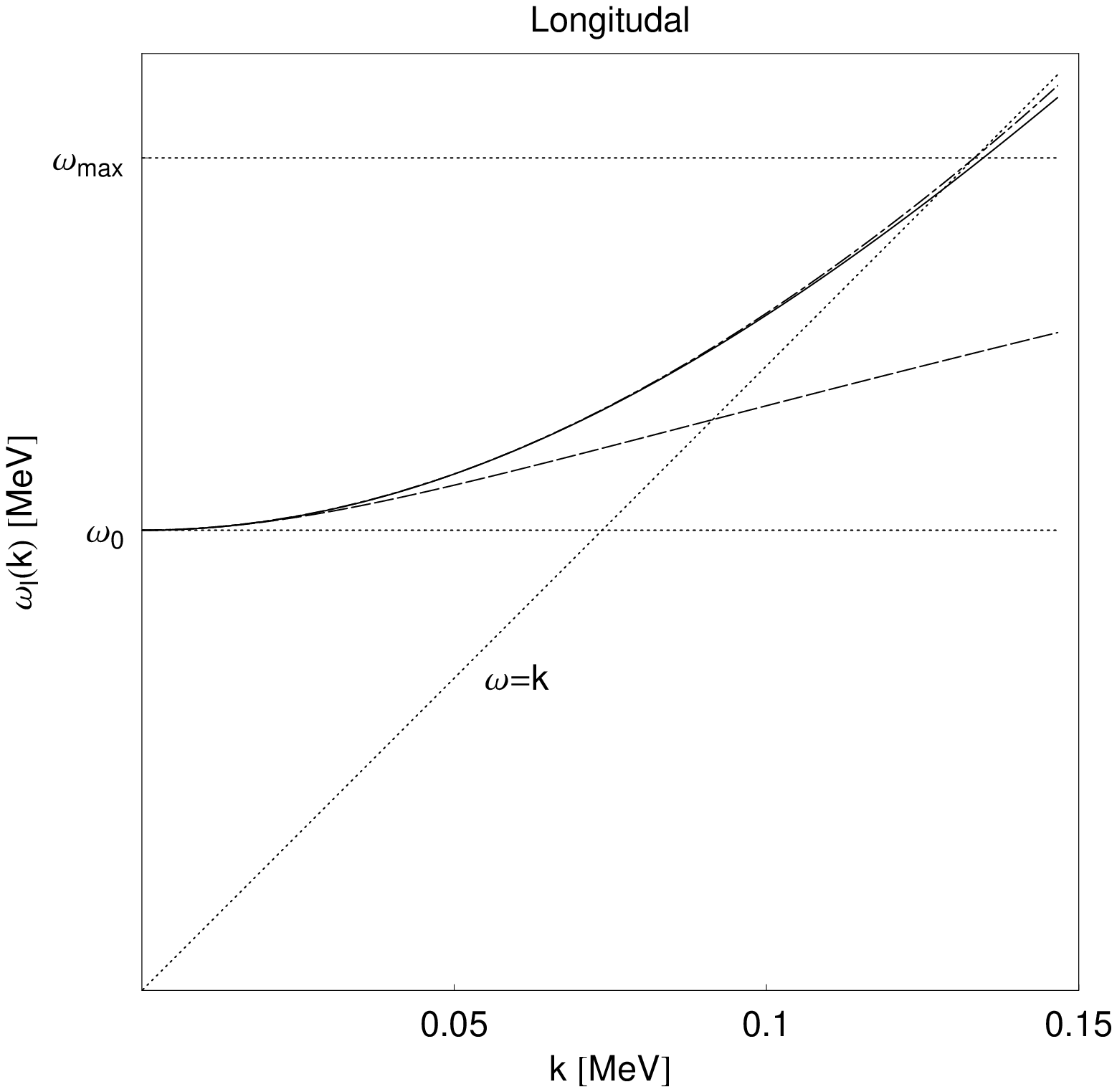}
  &
  \includegraphics[width=0.5\textwidth]{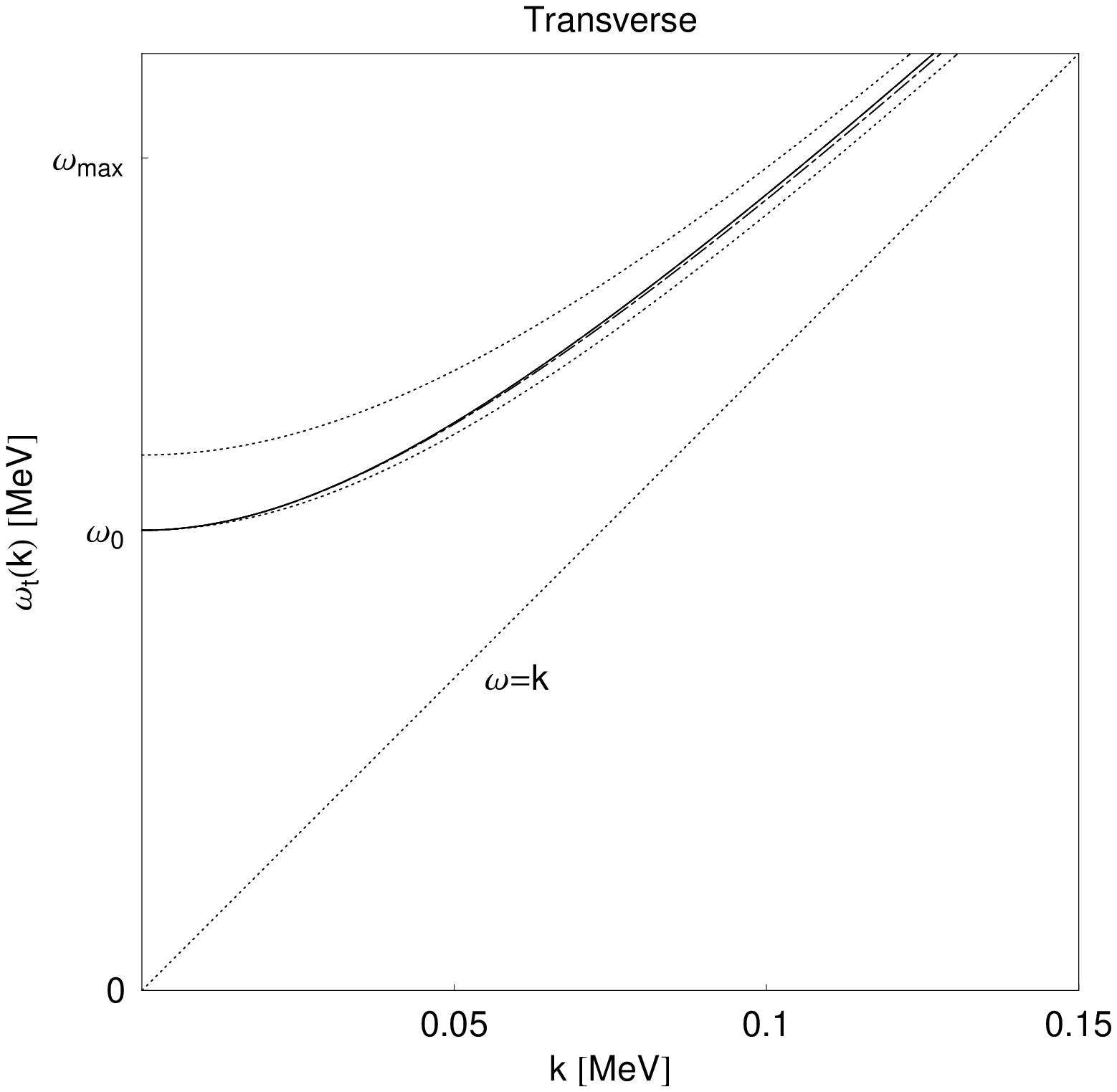}
\end{tabular}
\caption{\label{plasmon_dispersion} Longitudal and transverse 
plasmon dispersion relation $\omega_{l,t}(k)$ for 
plasma parameters from Table~\ref{tbl1}. Exact result (dot-sahed) is very close
to the Braaten \& Segel approximation (solid). Zero-order (dotted) and first order
 (dashed) approximations are very poor, especially for londitudal mode (left).}
\end{figure*}

\begin{figure*}
\begin{tabular}{cc}
  \includegraphics[width=0.5\textwidth]{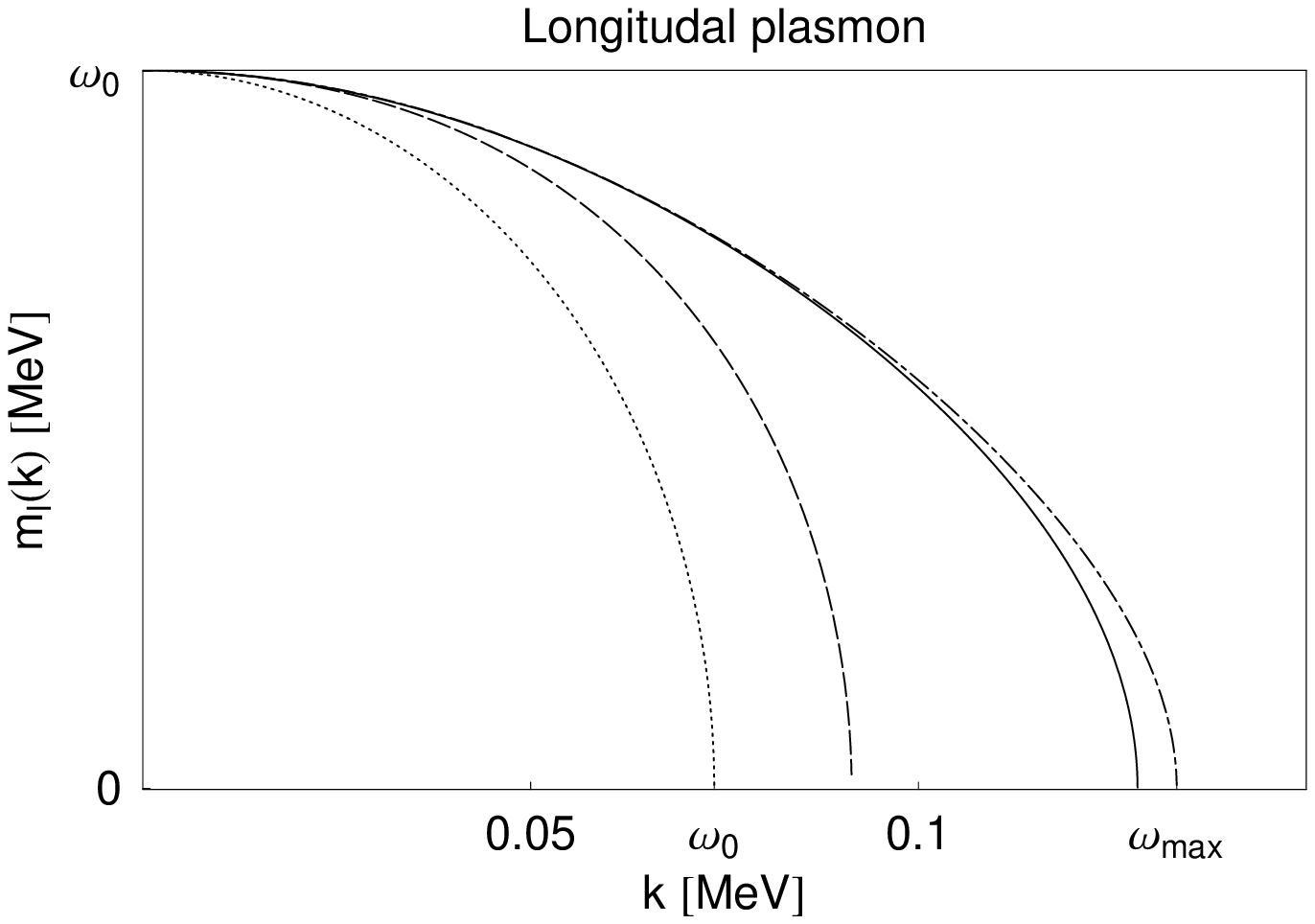}
  &
  \includegraphics[width=0.5\textwidth]{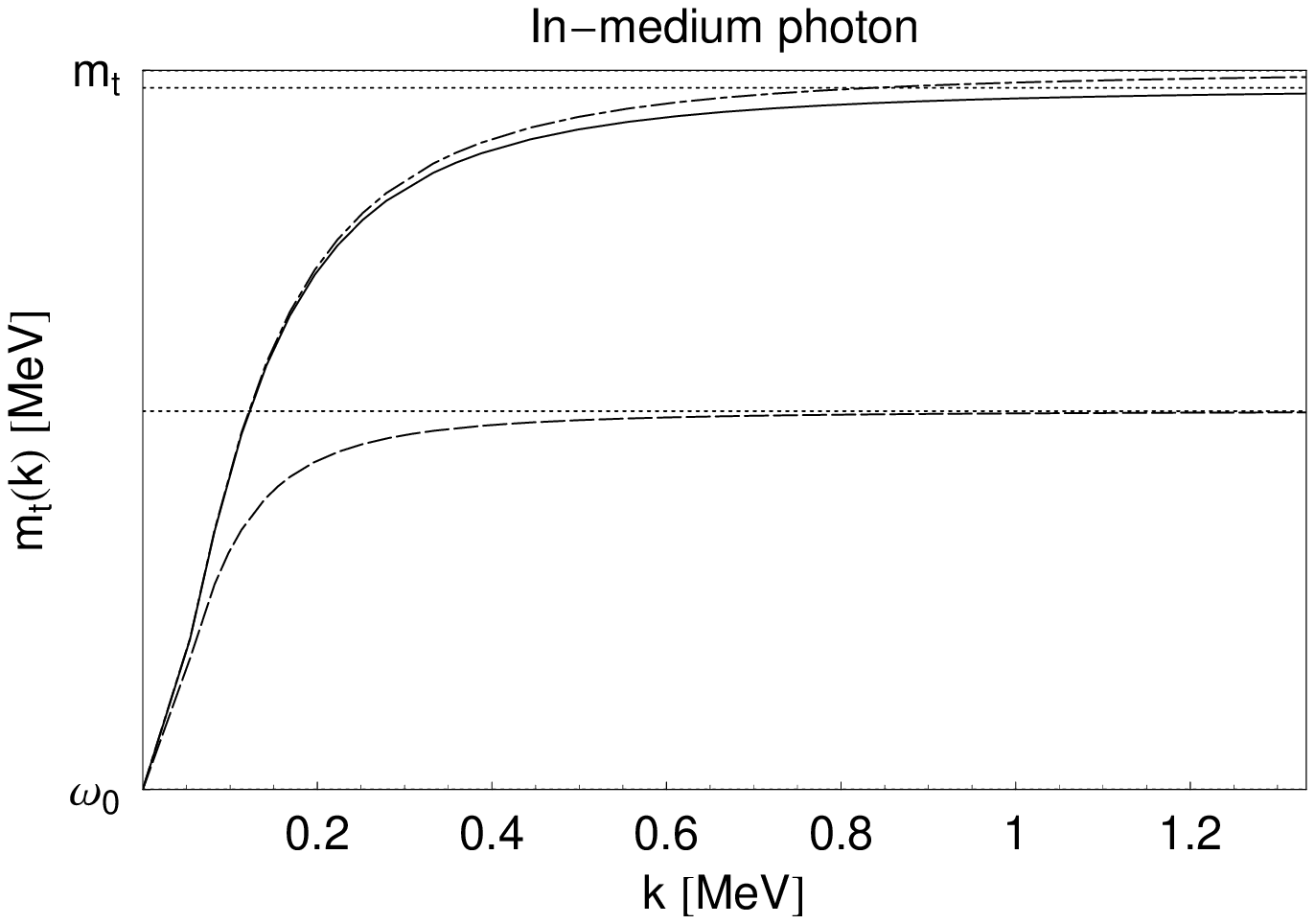}
\end{tabular}
\caption{\label{plasmon_mass} Longitudal and transverse  
plasmon mass. Dotted lines on the right panel show asymptotic transverse mass.
Line dashing the same as in Fig.~\ref{plasmon_dispersion}.}
\end{figure*}

Dispersion relations are solution to the equations \cite{BraatenSegel}:
\begin{subequations}
\label{disp_int}
\begin{equation}
\label{dispL_int}
k^2 =  \Pi_l \left ( \omega_l(k), k \right)
\end{equation}  
\begin{equation}
\label{dispT_int}
k^2 = \omega_t (k)^2 - \Pi_t \left ( \omega_t(k), k \right)
\end{equation}
\end{subequations}
where longitudal and transverse  polarization functions are given
as an integrals:
\begin{subequations}
\label{disp_intt}
\begin{equation}
\label{Pi_l}
\Pi_l= \frac{4 \alpha}{\pi} \int_0^\infty
\frac{p^2}{E} 
\left( 
\frac{\omega_l}{v k} \ln{\frac{\omega_l+ v k}{\omega_l - v k}} \!-\! 1
-\frac{\omega_l^2-k^2}{\omega_l^2 - v^2 k^2}
\right) (f_1+f_2)
\; dp.
\end{equation}  
\begin{equation}
\label{Pi_t}
\Pi_t = \frac{4 \alpha}{\pi} \int_0^\infty
\frac{p^2}{E} 
\left( 
\frac{\omega_t^2}{k^2} -  \frac{\omega_t^2-k^2}{k^2} \frac{\omega_t}{2 v k}
\ln{\frac{\omega_t+ v k}{\omega_t - v k}}
\right) (f_1+f_2) \; dp.
\end{equation}  
\end{subequations}

Typical example of the exact plasmon dispersion relations (dash-dotted) is presented in 
Fig.~\ref{plasmon_dispersion}.  
As solving eqns.~(\ref{dispL_int}, \ref{dispT_int}) with \eqref{disp_intt} 
is computationally intensive, three levels
of approximation for dispersion relations are widely used:
\begin{enumerate}
\item{zero-order analytical approximations}
\item{first order relativistic corrections}
\item{Braaten\&Segel approximation}
\end{enumerate}

\begin{subsubsection}{Approximations for longitudal plasmons}

For longitudal plasmons, the simplest zero-order approach
used in early calculations of Adams et al. \cite{Adams-Woo}
and more recently in \cite{Dutta} for photoneutrino process is to put simply:
\begin{equation}
\label{dispL_zero}
\omega(k) = \omega_0
\end{equation}
where $\omega_0$ is the plasma frequency \eqref{plasma_frequency}.
Maximum plasmon energy $\omega_{max}=\omega_0$ in this approximation.
Zero-order approximation is valid only for non-relativistic regime,
and leads to large errors of the total emissivity \cite{BPS}.

First relativistic correction to \eqref{dispL_zero}
has been introduced by Beaudet et al. \cite{BPS}. Dispersion relation $\omega_l(k)$
is given in an implicit form:
\begin{equation}
\label{dispL_1}
\omega_l^2 = \omega_0^2+ \frac{3}{5} \omega_1^2 \frac{k^2}{\omega_l^2},
\end{equation}
with maximum plasmon energy equal to:
\begin{equation}
\label{omega_max_1}
\omega_{max}^{(1)} = \sqrt{\omega_0^2 + \frac{3}{5} \omega_1^2}
\end{equation}
This approximation, however, do not introduce really serious
improvement (Figs.~\ref{plasmon_dispersion}, \ref{plasmon_mass} (left) \& \ref{plasmaL_fig}). Breaking point was publication 
of the Braaten\&Segel approximation \cite{BraatenSegel}. Using simple analytical equation:
\begin{equation}
\label{dispL_BS}
k^2 = 3 \, \frac{\omega_0^2}{v_{\ast}^2} 
\left (
\frac{\omega_l}{2 v_{\ast} k} \ln{\frac{\omega_l+v_{\ast} k}{\omega_l-v_{\ast} k}} -1
\right)
\end{equation}
where $v_\ast$ is defined in \eqref{electron_velocity} one is able to 
get almost exact dispersion relation, cf. 
Figs.~\ref{plasmon_dispersion} \& \ref{plasmon_mass}, left panels.
Solution to the eq.~\eqref{dispL_BS} exist in the range $1<k<k_{max}^{BS}$,
where, in this approximation, maximum longitudal plasmon momentum is:
\begin{equation}
\label{omega_max_BS}
  \left( \omega_{max}^{BS} \right)^2 =  
     \frac{3  \omega_0^2 }{2 v_{\ast}^2} \left( \frac{1}{2 v_\ast}
\ln{\frac{1+v_\ast}{1-v_\ast}} -1\right)
\end{equation}
what gives value slightly different than exact value 
(Fig.~\ref{plasmon_mass}, left), 
but required for consistency of the approximation.
\end{subsubsection}

\begin{subsubsection}{Approximations for transverse plasmons}

For photons in vacuum dispersion relation is $\omega_t=k$. Zero order
approximation for in-medium photons is:
\begin{subequations}
\label{dispT_zero}
\begin{equation}
\label{dispT_zero_1}
\omega_t^2 = \omega_0^2 + k^2, \quad k \ll \omega_0
\end{equation}
valid for small $k$ and:
\begin{equation}
\label{dispT_zero_2}
\omega_t^2 = m_t^2 + k^2, \quad k \gg \omega_0
\end{equation}
\end{subequations}
valid for very large $k$. Formulae \eqref{dispT_zero_1} and 
\eqref{dispT_zero_2} provide
lower and upper limit for realistic $\omega_t(k)$, respectively 
(cf.~Fig.~\ref{plasmon_dispersion}, right panel, dotted).
First order relativistic corrections lead to the formula:
\begin{equation}
\omega_t^2 = \omega_0^2 + k^2 + \frac{1}{5} \omega_1^2  \frac{k^2}{\omega_t^2}
\end{equation}
with asymptotic photon mass:
\begin{equation}
m_t^{(1)}  = \sqrt{\omega_0^2 + \omega_1^2/5}
\end{equation}

Finally, Braaten\&Segel approximation leads to:
\begin{equation}
\label{dispT_BS}
\omega_t^2 = k^2 + \omega_0^2 \frac{3\, \omega_t^2 }{2 \, v_\ast^2\, k^2}
\left(
1 - \frac{\omega_t^2- v_\ast^2 k^2}{2\, \omega_t\, v_\ast\, k}
\ln{\frac{\omega_t+v_{\ast} k}{\omega_t-v_{\ast} k}}
\right)
\end{equation}
Asymptotic photon mass $m_t^{BS}$ derived from \eqref{dispT_BS} is:
\begin{equation}
\label{tmass_BS}
\left( m_t^{BS} \right)^2 = \frac{3 \, \omega_0^2}{2 v_\ast^2}
\left (
1 - \frac{1-v_\ast^2}{2 v_\ast} \ln{\frac{1+v_\ast}{1-v_\ast}}
\right )
\end{equation}
This is slightly smaller (left panel of Fig.~\ref{plasmon_mass}, dashed) 
than exact value (solid line).

All four relations are presented in Fig.~\ref{plasmon_dispersion}.
Differences are clearly visible, but they are much less
pronounced for transverse than for longitudal plasmons. Inspection of Fig.~\ref{plasmon_mass}
reveals however, that in the large momentum regime asymptotic
behavior is correct only for exact integral relations \eqref{dispT_int} and
may be easily reproduced using \eqref{dispT_zero_2} with $m_t$ from \eqref{tmass}.

\end{subsubsection}

Let us recapitulate main conclusions. Braaten\&Segel approximation provide
reasonable approximation, as nonlinear equations \eqref{dispL_BS} and \eqref{dispT_BS} 
are easily solved using e.g. bisection method. Zero and first-order approximations 
(\ref{dispL_zero}, \ref{dispT_zero_1}, \ref{dispT_zero_2}) with limiting
values \eqref{sub-area} provide starting 
points and ranges. Approximation has been tested by \cite{Itoh_VIII}
and is considered as the best available \cite{Raffelt}. Errors for part of the  $kT$-$\mu$
plane where plasmaneutrino process {\em is not dominant} may be as large as 5\% \cite{Itoh_VIII}.
At present, these inaccuracies are irrelevant for any practical
application, and Braaten\&Segel approximation is recommended for all purposes.

\end{subsection}

\begin{subsection}{Plasmon decay rate}

In the Standard Model of electroweak interactions,
massive in-medium photons and longitudal plasmons
may decay into neutrino-antineutrino pairs:
\begin{equation}
\gamma^\ast \rightarrow \nu_x + \bar{\nu}_x.
\end{equation} 

In the first-order calculations two Feynmann diagrams
(Fig.~\ref{feynmann_diag}) contribute to 
decay rate \cite{BraatenSegel,Ratkovic}.

\begin{figure}
\begin{tabular}{cc}
\includegraphics[width=0.43\columnwidth]{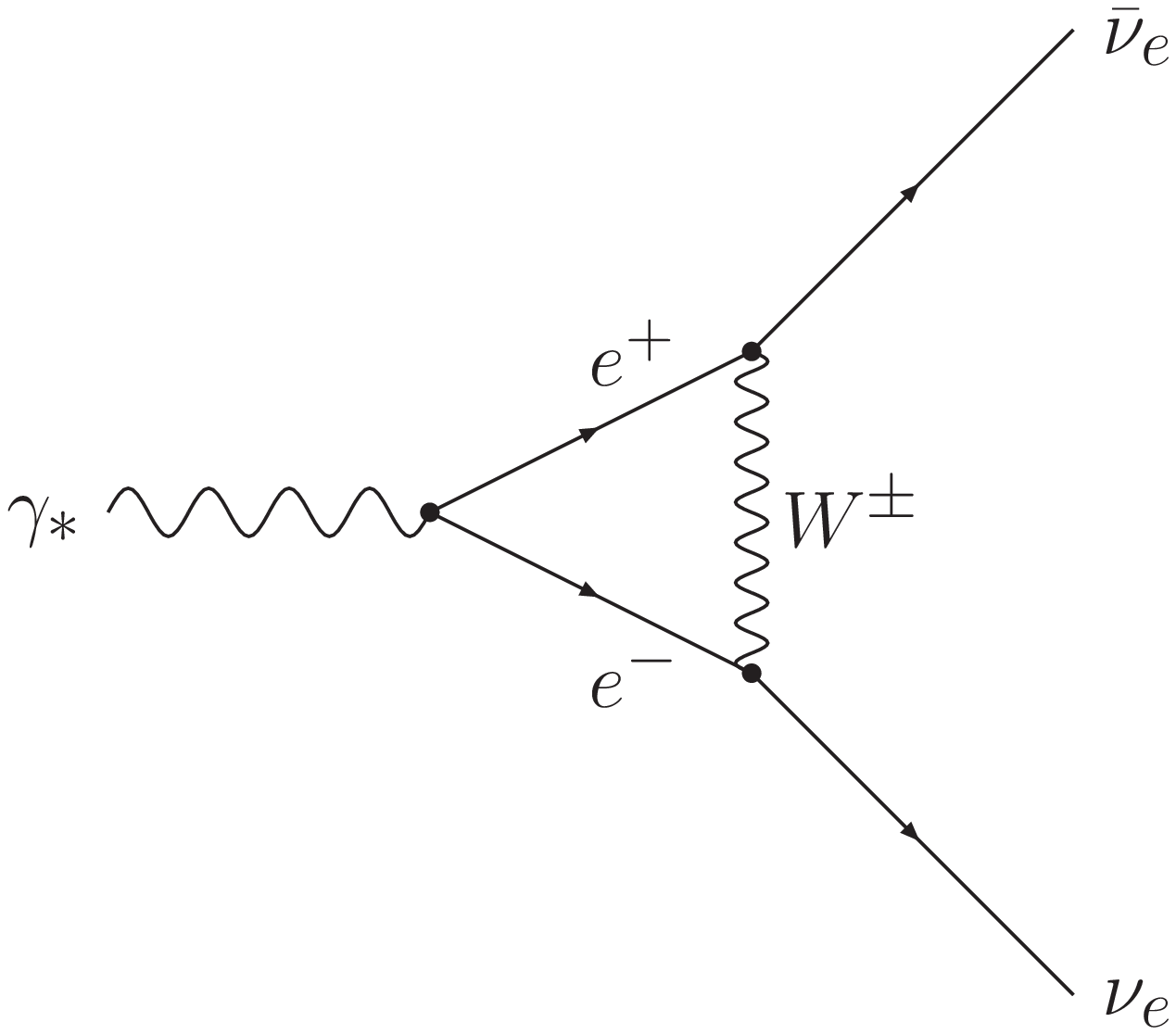}
&
\includegraphics[width=0.57\columnwidth]{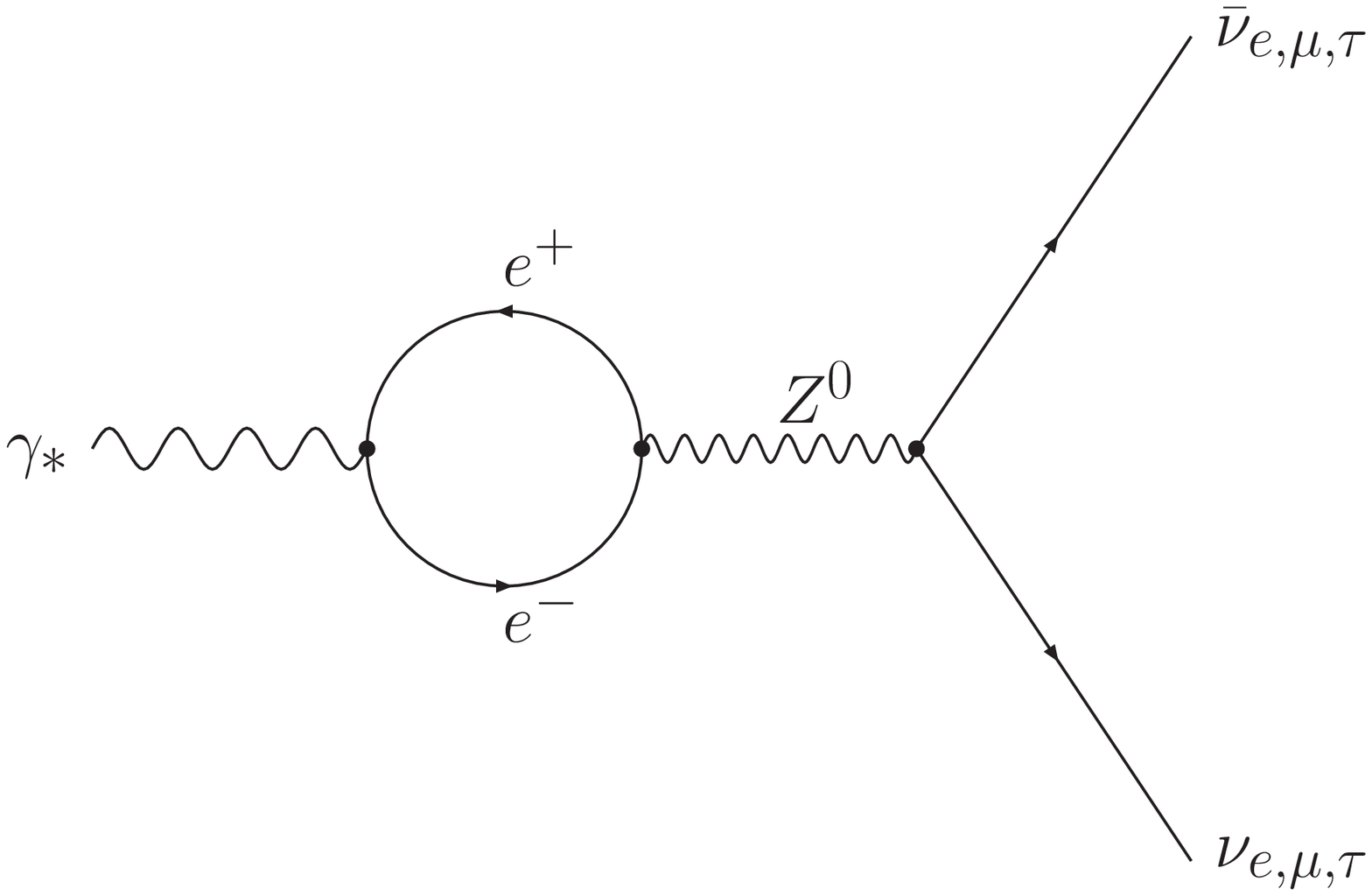}
\end{tabular}
\caption{\label{feynmann_diag} Fenmann diagrams for plasmon decay.}
\end{figure}

For the decay of the longitudal plasmon squared matrix element is:
\begin{subequations}
\begin{multline}
\label{Ml}
M_l^2 = \frac{G_F^2 C_V^2}{\pi  \alpha} \left( \omega_l^2-k^2 \right)^2
\;
\Biggl[
\frac{2 K \cdot Q_1 \; K \cdot Q_2}{K^2}+\\
\frac{2 \,\mathbf{k} \cdot \mathbf{q}_1 \; \mathbf{k} \cdot \mathbf{q}_2}{k^2}
-Q_1 \cdot Q_2
\Biggr] 
\end{multline}
where $K = (\omega,\mathbf{k})$ is four momentum
 of the plasmon. $Q_1 = (\mathcal{E}_1, \mathbf{q}_1)$  and 
$Q_2= (\mathcal{E}_2, \mathbf{q}_2)$ is four-momentum of the neutrino
and antineutrino, respectively.

Squared matrix element for decay of the massive photon is:
\begin{multline}
\label{Mt}
M_t^2 = \frac{G_F^2}{\pi  \alpha} 
\Biggl[
   \left(
    C_V^2 \Pi_t^2+
    C_A^2 \Pi_A^2 
    \right )
    \left  ( 
\mathcal{E}_1 \mathcal{E}_2 
- 
\frac{\mathbf{k} \cdot \mathbf{q}_1 \; \mathbf{k} \cdot \mathbf{q}_2}{k^2}
\right)  \\
+ 2 C_V C_A \Pi_t \Pi_A 
\frac{\mathcal{E}_1\; \mathbf{k} \cdot \mathbf{q}_2 - \mathcal{E}_2 \; \mathbf{k} \cdot \mathbf{q}_1
}{k} 
\Biggr]
\end{multline}
\end{subequations}
where $\Pi_t$ is defined in \eqref{Pi_t} and axial polarization function $\Pi_A$
reads:
\begin{equation}
\Pi_A = \frac{2\alpha}{ \pi} \frac{\omega_t^2\!-\!k^2}{k} \int_0^\infty
\frac{p^2}{E^2}
\left(
\frac{\omega_t}{2 v k} \ln{ \frac{\omega_t+v k}{\omega_t - v k} }
\!-\! \frac{\omega_t^2 - k^2}{\omega_t^2 - v^2 k^2}
\right)
 (f_1 - f_2)
\; dp
\end{equation}

Fermi constant is \mbox{$G_F/(\hbar c)^3 = 1.16637(1)\times 10^{-5} \, \mathrm{GeV}^{-2}$} 
\cite{PDBook} and, in standard model of electroweak interactions, vector and
axial coupling constants are: 
\begin{equation}
\label{CV_e}
C_V^e = \frac{1}{2} + 2 \sin^2{\theta_W}, \quad C_A^e = \frac{1}{2} 
\end{equation}
\begin{equation}
\label{CV_mu}
C_V^{\mu, \tau} = -\frac{1}{2} + 2 \sin^2{\theta_W}, \quad C_A^{\mu, \tau} = -\frac{1}{2}
\end{equation}
for electron and $\mu,\tau$ neutrinos, respectively. 
The Weinberg angle is $\sin^2{\theta_W}=0.23122(15)$ \cite{PDBook}.

Terms containing $C_A$ (so-called axial contribution) in \eqref{Mt} are frequently
treated separately \cite{Ratkovic} or removed at all \cite{Itoh_I}. 
In calculations concentrated on the total emissivity this is justified
as anti-symmetric term multiplied by $C_V C_A$ do not contribute at all and term
$C_A^2 \times  \ldots$ is suppressed relative to the term beginning with $C_V^2 \times \ldots$
by four orders of magnitude \cite{Itoh_I}. However, if one attempts
to compute neutrino energy spectrum all three terms should be
added together, as mixed V-A ,,channel'' alone  leads to negative emission
probability for some neutrino energy range (Fig.~\ref{VA_contrib}), 
what is physically unacceptable. 
These terms remains numerically small but {\em only} for
electron neutrinos.  For $\mu$ and $\tau$ neutrino spectra {\em axial}
part contributes at $\sim$ 1\% level due to very small 
value $C_V^{\mu,\tau} = -0.0376$ while still $C_A=-0.5$. ''Mixed'' term
leads to significant differences between $\nu_{\mu,\tau}$ and $\bar{\nu}_{\mu,\tau}$
spectra, cf.~Fig.~\ref{VA_contrib}. Relative contributions
of the three transverse ''channels'' for electron and $\mu,\tau$
are presented in Table~\ref{VA_coeff}.

\begin{table}
\caption{\label{VA_coeff} Relative weight of the $M_t^2$ \eqref{Mt} terms
for $e$ and $\mu,\tau$ neutrinos.}
\begin{tabular*}{\columnwidth}{c|ccc}
Flavor & Vector  & Axial & Mixed \\[3pt]
 & $ \frac{C_V^2 \omega_0^4}{(C_V \omega_0^2 + C_A \omega_A)^2}$ & $\frac{C_A^2 \omega_A^2}{(C_V \omega_0^2 + C_A \omega_A)^2}$ & $\quad \frac{2 C_V C_A \omega_0^2 \omega_A}{(C_V \omega_0^2 + C_A \omega_A)^2}$ \\
\hline
electron & 0.74 & 0.02 & 0.24 \\[3pt]
mu/tau   & 0.07 & 0.39 & 0.54 
\end{tabular*}
\end{table}

In general, all the terms in the 
squared matrix element \eqref{Mt} should be added. 
We have only {\em two} different spectra: longitudal and transverse one.

Particle production rate from plasma in thermal equilibrium is:
\begin{equation}
\label{rate}
R_i = \frac{g_i}{ (2 \pi)^5}
\int \; 
Z_i
\;
f_{\gamma^\ast}
\;
\delta^4 (K - Q_1 - Q_2)
\;
M_i^2
\,\frac{d^3 \mathbf{k} }{2 \omega_i}
\,
\frac{d^3 \mathbf{q}_1 }{2 \mathcal{E}_1}
\,
\frac{d^3 \mathbf{q}_2 }{2 \mathcal{E}_2}
\end{equation} 
where $i=l$ for longitudal mode and $i=t$ for transverse mode. 
Bose-Einstein distribution for plasmons $f_{\gamma^\ast}$ is: 
\begin{equation}
f_{\gamma^\ast} = \frac{1}{e^{\omega_{t,l}/kT} - 1}.
\end{equation}
and residue factors $Z_{t,l}$ are expressed by polarization functions $\Pi_{t,l}$ 
(\ref{Pi_t}, \ref{Pi_l}):
\begin{equation}
\label{res_t}
Z_t^{-1} = 1- \frac{\partial \Pi_t}{\partial \omega^2}
\end{equation}
\begin{equation}
\label{res_l}
Z_l^{-1} = -\frac{\omega_l^2}{k^2} \; \frac{\partial \Pi_l}{\omega^2}.
\end{equation}
For massive photons $g_t = 2$ and for longitudal plasmon $g_l = 1$.

Differential rates\footnote{
\label{DDrate2}
Double differential rate $d^2 R_i/d\mathcal{E}d\cos{\theta}$ has an identical
form as \eqref{DiffRate} but now four momenta cannot be given explicitly, unless
simple analytical approximation for $\omega_i(k)$ is used. Analytical
approximations for the specrum shape are derived this way.
}
has been derived for the first time  in \cite{Ratkovic}.
Here, we present result in the form valid for both types of plasmons, ready
for calculations using any available form of dispersion relation:
\begin{equation}
\label{DiffRate}
\frac{d^2 R_i}{d \mathcal{E}_1 \, d \mathcal{E}_2} = 
\frac{g_i}{\pi^4} \, Z_i M_i^2
f_{\gamma^\ast}\, J_i \; \mathcal{S}
\end{equation}
where $i=l$ or $i=t$.
Product $\mathcal{S}$ of the unit step functions $\Theta$ in \eqref{DiffRate} restrict
result to the kinematically allowed area:
\begin{equation}
\label{int_area}
\mathcal{S} =
\Theta(4 \mathcal{E}_1 \mathcal{E}_2 - m_i^2) 
\Theta(\mathcal{E}_1+\mathcal{E}_2 -\omega_0)
\Theta(\omega_{max}-\mathcal{E}_1-\mathcal{E}_2) 
\end{equation}

Four-momenta in the squared matrix element are:
\begin{eqnarray*}
Q_1 &=& (\mathcal{E}_1, 0 , 0,\mathcal{E}_1 ) \\
Q_1 &=& (\mathcal{E}_2, \mathcal{E}_2 \sin{\theta}, 0 , \mathcal{E}_2 \cos{\theta}) \\
K &=& (\mathcal{E}_1+\mathcal{E}_2, \mathcal{E}_2 \sin{\theta}, 0, \mathcal{E}_1+  \mathcal{E}_2 \cos{\theta})\\
m_i^2 &=& K \cdot K = (\mathcal{E}_1 + \mathcal{E}_2)^2-k'^2   \\
\cos{\theta} &=& \frac{{k'}^2-\mathcal{E}_1^2-\mathcal{E}_2^2}{2 \mathcal{E}_1 \mathcal{E}_2} \\
k'  &=& \omega_{l,t}^{-1}(\mathcal{E}_1+\mathcal{E}_2)    \\
\omega_i &=&    \mathcal{E}_1+\mathcal{E}_2
\end{eqnarray*}
where $\omega_i^{-1}$ denotes function {\em inverse} to the dispersion relation.
Jacobian $J_i$ arising from Dirac delta integration in \eqref{rate} is:
\begin{equation}
J_i^{-1} = \frac{\mathcal{E}_1 \mathcal{E}_2}{k'} 
\frac{\partial \omega_i}{\partial k} \Bigg|_{k=k'}.
\end{equation}
Residue factors $Z_i$ are given in \eqref{res_l} and \eqref{res_t}. Maximum
energy $\omega_{max}$ in \eqref{int_area} for longitudal plasmons must be in the
agreement with particular approximation used for $\omega_l(k)$:
$\omega_0$, \eqref{omega_max_1} or \eqref{omega_max_BS}
for zero-order \eqref{dispL_zero}, first-order \eqref{dispL_1} or Braaten\&Segel \eqref{dispL_BS}
approximation, respectively. For transverse plasmons $\omega_{max} \to \infty$
and last $\Theta$ function in \eqref{int_area} has no effect and may be omitted.

\end{subsection}

\begin{subsection}{Longitudal neutrino spectrum}

\begin{subsubsection}{Analytical approximation}

We begin with general remark on the spectrum.
Note, that eq.~\eqref{DiffRate} is symmetric for longitudal mode 
under change $\mathcal{E}_{1,2} \to \mathcal{E}_{2,1}$ because \eqref{Ml}
is symmetric with respect to exchange $Q_{1,2} \to Q_{2,1}$. 
Resulting energy spectrum is thus identical for neutrinos and antineutrinos.
This is not true for transverse plasmons with axial
contribution included, cf.~Sect.~\ref{transverse_spectrum}.

Using zero-order dispersion relation for longitudal plasmons
\eqref{dispL_zero} we are able to express spectrum by the elementary functions.
Longitudal residue factor $Z_t$ is now:
\begin{equation}
Z_l^0 = 1,
\end{equation}
and Jacobian $J_l$ resulting from the integration 
of the Dirac delta function is:
\begin{equation}
J_l^0 = 1. 
\end{equation}

Now, differential rate $d^2R/d\mathcal{E}d\cos{\theta}$  (cf. \eqref{DiffRate}
and footnote \ref{DDrate2}) becomes much more simple
and integral over $d \cos{\theta}$ can be evaluated analytically.
Finally, we get the longitudal spectrum:
\begin{equation}
\label{widmoLzero}
\frac{d R}{d \mathcal{E} } \equiv \lambda( \mathcal{E} ) = 
\frac{ {G_F}^2 \, {C_V}^2 \, {\omega_0}^7}{1260 \, \pi^4\, \alpha\, \hbar^3\, c^9} \;
\frac{
f(\mathcal{E}/\omega_0)
}
{
e^{\omega_0/kT}-1
}
\end{equation}

where normalized spectrum is:
\begin{eqnarray}
\nonumber
f(x)& =& \frac{105}{32} \; \Bigl[ 4 x (x-1) (8 x^4 -16 x^3 + 2 x^2  + 6 x -3)  \\ 
&+& 
3 (1-2x)^2 \ln (1-2x)^2 \Bigr] 
\end{eqnarray} 
Let us note that $f$ is undefined at $x=1/2$; use limit instead:
$$
\lim_{x \to 1/2} f(x) = 105/32.
$$
Function $f(x)$  is symmetric with respect to point $x=1/2$, where $f$
has a maximum value (Fig.~\ref{plasmaL_fig}, dotted line).

In this limit, correct for non-relativistic, non-degenerate plasma, 
average neutrino and antineutrino energy is $\langle \mathcal{E} \rangle = \omega_0/2$
and maximum $\nu$ energy is $\omega_0$. 

Inspection of Fig.~\ref{plasmaL_fig}
reveals little difference between analytical result \eqref{widmoLzero}
and result obtained with first-order relativistic corrections to the
dispersion relation \eqref{dispL_1}.
\end{subsubsection}

\begin{figure}
\includegraphics[width=\columnwidth]{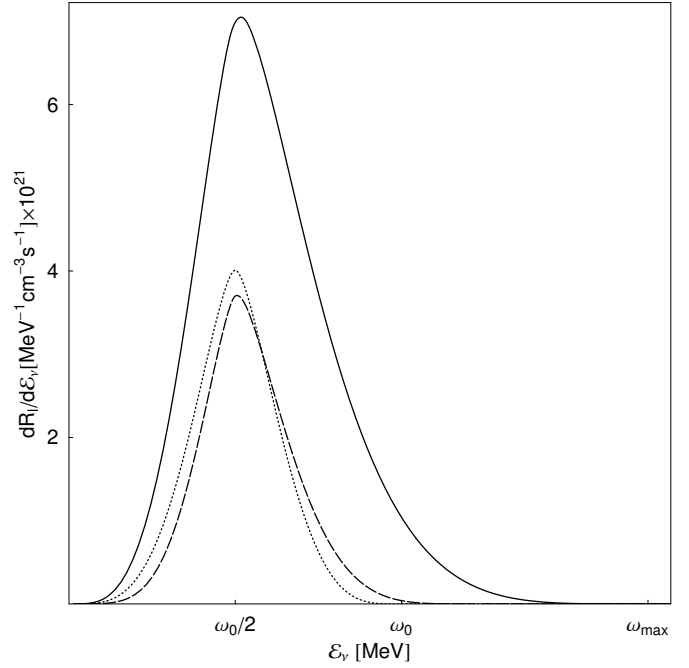}
\caption{\label{plasmaL_fig} Longitudal plasmon approximate analytical \eqref{widmoLzero}
neutrino spectrum (dotted), with first-order correction used by BPS \cite{BPS}
(dashed), and spectrum computed using \cite{BraatenSegel} dispersion relation
(solid). Plasma properties according to Table~\ref{tbl1}.
}
\end{figure}

\begin{subsubsection}{Numerical results}

Simple formula \eqref{widmoLzero} significantly underestimates flux and the 
maximum neutrino energy, equal to $\omega_{max}$ rather than $\omega_0$.
Therefore we have used Braaten \& Segel approximation for longitudal plasmon 
dispersion relation.

To derive spectrum we will use form of differential rate \eqref{DiffRate}
provided by \cite{Ratkovic}.
In the Braaten\&Segel approximation:
$$
Z_l^{BS} = \frac{\omega_l^2}{\omega_l^2-k^2} \;  
\frac{2 (\omega_l^2-v_{\ast}^2 k^2)}{3 \omega_0^2-\omega_l^2+v_{\ast}^2 k^2}
,
$$
$$
J_l^{BS}=\left |
\frac{k^2}{\mathcal{E}_1 \mathcal{E}_2}
\frac{1-\beta_l}{\omega_l \beta_l}
\right |, 
$$
$$
\beta_l^{BS} = \frac{3 \omega_0^2}{2 v_\ast^3 }
\left(
\frac{3 \omega_l}{2 k^3} \ln{\frac{\omega_l+ v_\ast k}{\omega_l - v_\ast k}}
- \frac{\omega_l^2 v_\ast}{k^2 (\omega_l^2-v_\ast^2 k^2)} - \frac{2 v_\ast}{k^2}
\right)   .
$$
Spectrum is computed as an integral of \eqref{DiffRate}
over $d\mathcal{E}_2$.
Example result is presented in Fig.~\ref{plasmaL_fig}. Integration of 
the function in Fig.~\ref{plasmaL_fig} over neutrino
energy gives result in well agreement with both (30) from \cite{BraatenSegel} 
and (54) from \cite{Ratkovic}.  

\end{subsubsection}

\end{subsection}

\begin{subsection}{Transverse plasmon decay spectrum \label{transverse_spectrum}}

\begin{subsubsection}{Analytical approximation}

Derivation of massive in-medium photon decay spectrum closely follows
previous subsection. Semi-analytical formula can be derived for dispersion relations
\eqref{dispT_zero}.
For dispersion relation \eqref{dispT_zero_2} 
transverse residue factor $Z_t$ is:
\begin{equation}
Z_t^0 = 1,
\end{equation}
polarization function $\Pi_t$ is equal to:
\begin{equation}
\Pi_t^0 = m_t^2,
\end{equation}
and Jacobian resulting from integration 
of the Dirac delta function $J_t$ is:
\begin{equation}
J_t^0 = \frac{\mathcal{E}_1+\mathcal{E}_2}{\mathcal{E}_1 \mathcal{E}_2}.
\end{equation}

\begin{figure}
\includegraphics[width=\columnwidth]{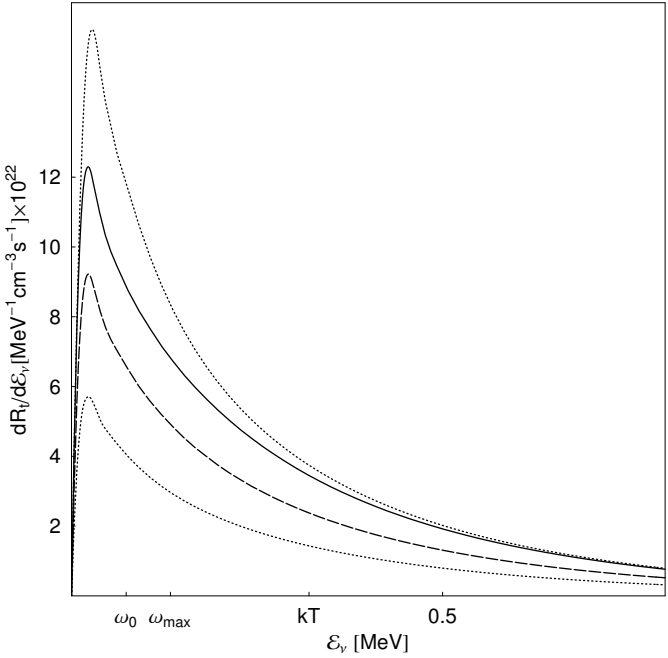}
\caption{
\label{plasmaT_fig}
Transverse plasmaneutrino spectrum computed from \cite{BraatenSegel} 
approximation (solid) with upper \eqref{dispT_zero_2} and 
lower \eqref{dispT_zero_1} limits for the dispersion relation
(dotted). First-order relativistic correction
leads to the spectrum shown as dashed line. Plasma parameters as in Fig.~\ref{plasmaL_fig}.
}
\end{figure}

Approximate spectrum, neglecting differences between
neutrinos and antineutrinos, is given by the following integral:
\begin{equation}
\label{widmoTzero}
\lambda(\mathcal{E}) = \frac{G_F^2 C_V^2}{64 \, \pi^4 \alpha} 
\frac{m_t^7}{\hbar^3 c^9}
\int_{-1}^1 \frac{P( \cos{\theta}, \mathcal{E}/m_t) \; d\cos{\theta} }
{\exp \left[ (\mathcal{E}+\frac{m_t^2}{2 \mathcal{E} (1-\cos{\theta})})/kT \right]-1}
\end{equation}
where rational function $P(ct,x)$ is:
\begin{equation}
\label{Pfunct}
P = \frac{1+2 (ct-1)^2 (2 x^2 -1) x^2}
{ x (ct-1)^2 [ 1- 2 ct (ct-1) x^2 +2 (ct-1)^2) x^4 ]}
\end{equation}

Result presented in Fig.~\ref{plasmaT_fig} show
that spectrum \eqref{widmoTzero} obtained with dispersion relation \eqref{dispT_zero_2}
agree well in both low and high neutrino energy part
with spectrum obtained from Braaten\&Segel approximation for
dispersion relations. Dispersion relation \eqref{dispT_zero_1}
produces much larger error, and spectrum {\em nowhere} agree with correct
result. This fact is not a big surprise: as was pointed out by Braaten \cite{BraatenPRL}
dispersion relation is crucial.
Therefore, all previous results, including seminal BPS work \cite{BPS},
could be easily improved just by the trivial replacement
$\omega_0 \to m_t$. Moreover, closely related photoneutrino
process also has been computed \cite{BPS,Itoh_I,Schinder,Dicus} 
with simplified dispersion relation
\eqref{dispT_zero_1} with $\omega_0$. 
One exception is work of Esposito et. al. \cite{Esposito}. It remains unclear however,
which result is better, as accurate dispersion relations have never
been used within photoneutrino process context. For plasmaneutrino,
Eq.~\eqref{dispT_zero_2} is much better approximation than \eqref{dispT_zero_1},
especially if one put $m_t$ from exact formula \eqref{tmass}.     
High energy tail of the spectrum also will be exact in this case.

As formula \eqref{widmoTzero} agree perfectly with the tail
of the spectrum, we may use it to derive very useful analytical expression.
Leaving only leading terms of the rational function \eqref{Pfunct}
$$
P(ct,x) \sim x^{-1} (1-ct)^{-2}
$$ 
one is able to compute integral
\eqref{widmoTzero} analytically:
\begin{equation}
\label{widmoTanal}
\lambda(\mathcal{E}) \simeq  \frac{G_F^2 C_V^2}{64 \, \pi^4 \alpha} 
\frac{m_t^6}{\hbar^3 c^9} \left [
\kappa      - \frac{2}{a} \ln{\left( e^{a \kappa/2} -1\right)}
\right]
\end{equation}
where $\kappa=2x + (2x)^{-1}$, $x=\mathcal{E}/m_t$, $a=m_t/kT$.
Interestingly, spectrum \eqref{widmoTanal} is invariant
under transformation:
$$
\mathcal{E}' \mathcal{E} = m_t^2/4
$$
and all results obtained for high energy tail of the spectrum
immediately may be transformed for low-energy approximation. 
The asymptotic behavior of \eqref{widmoTanal}  for $\mathcal{E} \gg kT$
is of main interest:
\begin{equation}
\label{widmoTtail}
\lambda(\mathcal{E}_\nu)  
= A\; kT\, m_t^6 \,   \exp{ \left(-\frac{\mathcal{E}_\nu}{kT} \right)}
\end{equation}
where for electron neutrinos : 
$$
A= \frac{G_F^2 C_V^2}{8 \pi^4 \alpha} 
\frac{1}{\hbar^4 c^9} = 2.115 \times 10^{30} \; 
[\mathrm{MeV}^{-8} \mathrm{cm}^{-3} \mathrm{s}^{-1}] 
$$
and $m_t$, $kT$ are in MeV. For $\mu, \tau$ neutrinos just
replace $A$ with $A\,(C_V^{\mu,\tau}/C_V^e)^2$.

Formula \eqref{widmoTtail} gives also quite reasonable estimates
of the total emissivity $Q_t$ and mean neutrino energies $\langle \mathcal{E}_\nu \rangle$:
\begin{subequations}
\begin{equation}
Q_t = A\; kT^3\, m_t^6 
\end{equation}
\begin{equation}
\langle \mathcal{E}_\nu \rangle = kT
\end{equation}
\end{subequations}
For a comparison, Braaten \& Segel \cite{BraatenSegel} derived exact formulae
in the high temperature limit $kT \gg \omega_0$:
\begin{subequations}
\begin{equation}
Q_t^{BS} = \frac{G_F^2 C_V^2 \zeta(3) }{12 \pi^4 \alpha}\; kT^3\, m_t^6  = 0.8\, A\; kT^3\, m_t^6
\end{equation}
\begin{equation}
\langle \mathcal{E}_\nu^{BS} \rangle = \frac{6 \zeta(3)}{\pi^2} kT = 0.73\, kT
\end{equation}
\end{subequations}
Formulae above agree with $\sim$25\% error in the leading coefficients.

\end{subsubsection}

\begin{subsubsection}{Numerical results}
Calculation of the spectrum in the framework of Braaten\&Segel
approximation requires residue factor, polarization function 
\cite{BraatenSegel} (transverse\&axial) and Jacobian \cite{Ratkovic}:
\begin{equation}
Z_t^{BS} =  \frac{2\, \omega_t^2 \, (\omega_t^2 - v_\ast^2 \, k^2)}
{3\, \omega_0^2 \omega_t^2 + (\omega_t^2 + k^2)(\omega_t^2 - v_\ast^2 \, k^2)
-2 \, \omega_t^2 (\omega_t^2 - k^2)
},
\end{equation}
\begin{equation}
\Pi_t^{BS} = \frac{3 \, \omega_0^2}{2 v_\ast^2}
\left( 
\frac{\omega_t^2}{k^2} - \frac{\omega_t^2-v_\ast^2 k^2}{k^2} \, \frac{\omega_t}{2 v_\ast k}
\ln{\frac{\omega_t+ v_\ast k}{\omega_t - v_\ast k}}
\right),
\end{equation}
\begin{equation}
\Pi_A^{BS} = \omega_A\,k\; \frac{\omega_t^2 - k^2}{\omega_t^2 - v_\ast^2 k^2} 
\;
\frac{3\, \omega_0^2 - 2\, (\omega_t^2 -k^2)}{\omega_0^2},
\end{equation}
\begin{equation}
J_t^{BS} = \frac{\mathcal{E}_1 + \mathcal{E}_2}{\mathcal{E}_1 \mathcal{E}_2} 
\left| \frac{1-\beta_t^{BS} }{1- \frac{\omega_t^2}{k^2} \beta_t^{BS}} \right|
\end{equation}
\begin{equation}
\beta_t^{BS} = \frac{9 \omega_0^2}{4 v_\ast^2 k^2} 
\left [
1+ \frac{1}{6} \left(  \frac{v_\ast k}{\omega_t} - \frac{3 \omega_t}{v_\ast k} \right)
\ln{\frac{\omega_t+ v_\ast k}{\omega_t - v_\ast k}}
\right]
\end{equation}

Example spectrum, computed as an integral of \eqref{DiffRate} over $d \mathcal{E}_2$ 
is shown in Fig.~\ref{plasmaT_fig}.

\end{subsubsection}

\begin{figure}
\includegraphics[width=\columnwidth]{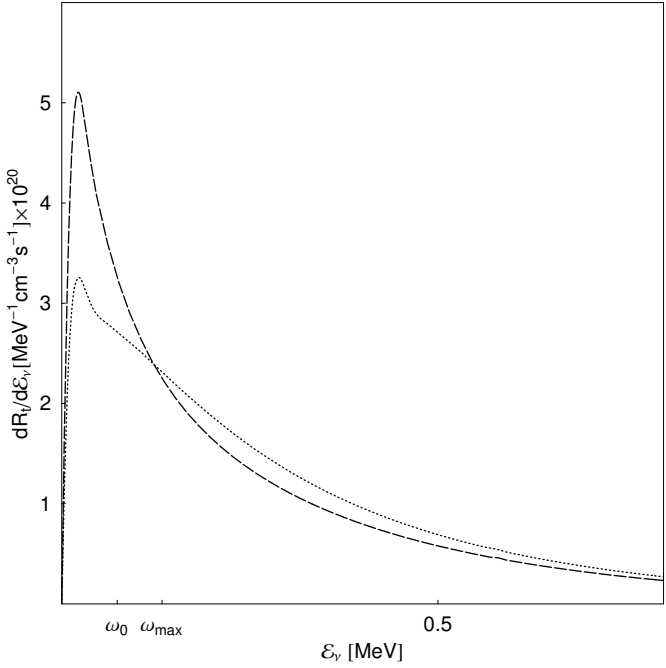}
\caption{\label{VA_contrib} Spectrum of the muon
neutrinos (dotted) and antineutrinos (dashed) from transverse
plasmon decay.
Contributions to the spectra from so-called
 mixed ,,vector-axial channel'' produces significant differences. 
For electron flavor, contribution from ''mixed channel'' lead
to unimportant differences. For both flavors contribution
from ''axial channel'' remains relatively small: $10^{-4}$ for $\nu_e$
and $10^{-2}$ for $\nu_\mu$. Overall contribution to the total
emissivity from $\mu,\tau$ flavors is suppressed relatively
to electron flavor by a factor $(C_V^{\mu,\tau}/C_V^e)^2 \simeq 3.3 \times 10^{-3}$.  
}
\end{figure}

\end{subsection}

\end{section}

\begin{figure}
\includegraphics[width=\columnwidth]{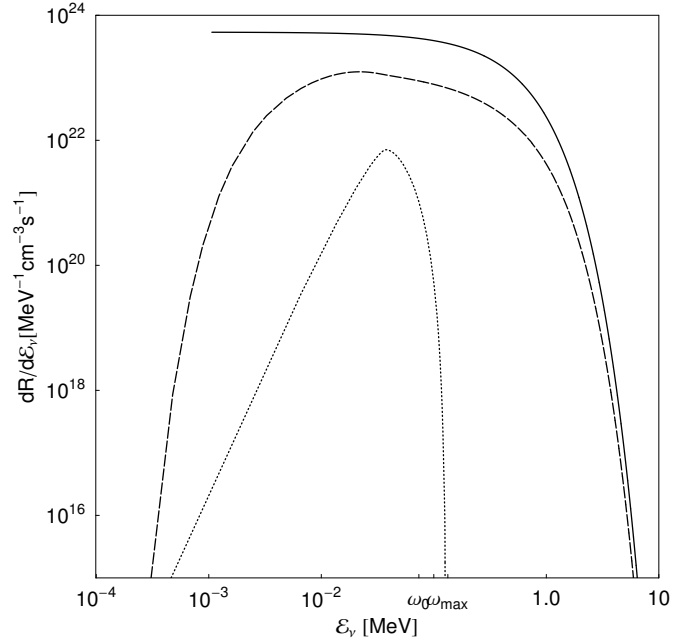}
\caption{
Typical spectra from the plasma process. Dotted line
is a longitudal and dashed transverse spectrum.
Only $\sim \exp(-\mathcal{E}_{\nu}/kT)$ tail of the transverse spectrum 
(solid line) contributes to (possibly) detectable signal. Plasma properties
according to Table~\ref{tbl1}.
}
\end{figure}

\section{Summary}

Main new results presented in the article are analytical formulae for neutrino spectra
(\ref{widmoLzero}, \ref{widmoTzero}) and exact analytical formula \eqref{widmoTtail} for the
high energy tail of the transverse spectrum. The latter is of main interest from
the detection of astrophysical sources point of view:
recently available detection techniques are unable to detect keV plasmaneutrinos
emitted with typical energies $\langle \mathcal{E}_{\nu} \rangle \sim \omega_0/2$ 
(Fig.~\ref{plasmaL_fig}, \ref{plasmaT_fig}), where
$\omega_0$ is the plasma frequency \eqref{plasma_frequency}.
Tail behavior of the transverse spectrum quickly ''decouple'' from $\omega_0$ dominated
maximum area, and becomes dominated by temperature-dependent 
term $\exp{(-\mathcal{E}_\nu/kT)}$. Calculation of the events
in the detector is then straightforward, as detector threshold
in the realistic experiment will be above maximum area. This approach
is much more reliable compared to the typical practice, where
an average neutrino energy is used as a parameter in an arbitrary
analytical formula.

Analytical formulae for the spectrum are shown to be a poor approximation
of the realistic situation, especially for longitudal plasmons (Fig.~\ref{plasmaL_fig}).
This is in the agreement with general remarks on the dispersion relations
presented by Braaten \cite{BraatenPRL}. On the contrary, 
Braaten \& Segel \cite{BraatenSegel}
approximation is shown to be a very good approach not only for the total emissivities,
but also for the spectrum. Exception is the tail of the massive
photon decay neutrino spectrum: Braaten \& Segel \cite{BraatenSegel} formulae lead to
underestimate of the thermal photon mass while the formula \eqref{widmoTtail}
gives exact result. Numerical difference between $m_t$ from \eqref{tmass} and \eqref{tmass_BS}
is however small \cite{BraatenSegel}. Calculating of the emissivities
by the spectrum integration seems much longer route compared to typical
methods, but we are given much more insight into process details. 
For example, we obtain exact formula for the tail for free this way.
Interesting surprise revealed in the course of our calculations is importance
of the high-momentum behavior of the massive photon. While mathematically
identical to simplest approach used in the early calculations, formula \eqref{dispT_zero_2}
gives much better approximation for the total emissivity than \eqref{dispT_zero_1}.

\begin{acknowledgement}
This work was supported by grant of Polish  Ministry of Education and Science (former
Ministry of Scientific Research and Information Technology, 
now Ministry of Science and Higher Education) No. 1~P03D~005~28.
\end{acknowledgement}


\begin{thebibliography}{28}

\bibitem{Arnett}
D.~{Arnett}, \emph{{Supernovae and nucleosynthesis}} (Princeton University
  Press, 1996)

\bibitem{Bisnovaty}
G.S. {Bisnovatyi-Kogan}, \emph{{Stellar physics. Vol.1: Fundamental concepts
  and stellar equilibrium}} (Springer, 2001)

\bibitem{Itoh_I}
H.~{Munakata}, Y.~{Kohyama}, N.~{Itoh}, Astrophys. J \textbf{296}, 197 (1985)

\bibitem{Itoh_I_erratum}
H.~{Munakata}, Y.~{Kohyama}, N.~{Itoh}, Astrophys. J \textbf{304}, 580 (1986)

\bibitem{Itoh_II}
Y.~{Kohyama}, N.~{Itoh}, H.~{Munakata}, Astrophys. J \textbf{310}, 815 (1986)

\bibitem{Itoh_III}
N.~{Itoh}, T.~{Adachi}, M.~{Nakagawa}, Y.~{Kohyama}, H.~{Munakata}, Astrophys. J
  \textbf{339}, 354 (1989)

\bibitem{Itoh_III_erratum}
N.~{Itoh}, T.~{Adachi}, M.~{Nakagawa}, Y.~{Kohyama}, H.~{Munakata}, Astrophys. J
  \textbf{360}, 741 (1990)

\bibitem{Itoh_IV}
N.~{Itoh}, H.~{Mutoh}, A.~{Hikita}, Y.~{Kohyama}, Astrophys. J \textbf{395}, 622 (1992)

\bibitem{Itoh_V}
Y.~{Kohyama}, N.~{Itoh}, A.~{Obama}, H.~{Mutoh}, Astrophys. J \textbf{415}, 267 (1993)

\bibitem{Itoh_VI}
Y.~{Kohyama}, N.~{Itoh}, A.~{Obama}, H.~{Hayashi}, Astrophys. J \textbf{431}, 761 (1994)

\bibitem{Itoh_VII}
N.~{Itoh}, H.~{Hayashi}, A.~{Nishikawa}, Y.~{Kohyama}, Astrophys. Js \textbf{102}, 411
  (1996)

\bibitem{BPS}
G.~{Beaudet}, V.~{Petrosian}, E.E. {Salpeter}, Astrophys. J \textbf{150}, 979 (1967)

\bibitem{Adams-Woo}
J.B. {Adams}, M.A. {Ruderman}, C.H. {Woo}, Physical Review \textbf{129}, 1383
  (1963)

\bibitem{Dicus}
D.A. {Dicus}, Phys. Rev. D \textbf{6}, 941 (1972)

\bibitem{BraatenSegel}
E.~Braaten, D.~Segel, Phys. Rev. D \textbf{48}(4), 1478 (1993)

\bibitem{BraatenPRL}
E.~Braaten, Phys. Rev. Lett. \textbf{66}(13), 1655 (1991)

\bibitem{Schinder}
P.J. Schinder, D.N. Schramm, P.J. Wiita, S.H. Margolis, D.L. Tubbs, 
Astrophys. J \textbf{313}, 531 (1987)

\bibitem{BlinnikovRudzskij}
S.I. {Blinnikov}, M.A. {Rudzskij}, Astron. Zh. \textbf{66}, 730 (1989)

\bibitem{BlinnikovRudzskij2}
S.I. {Blinnikov}, M.A. {Rudzskii}, Sov. Astron. \textbf{33}, 377 (1989)

\bibitem{Raffelt}
M.~{Haft}, G.~{Raffelt}, A.~{Weiss}, Astrophys. J \textbf{425}, 222 (1994)

\bibitem{ReinesCowan}
F.~Reines, C.L. Cowan, Phys. Rev. \textbf{113}(1), 273 (1959)

\bibitem{SK}
\verb1 http://www-sk.icrr.u-tokyo.ac.jp/sk/index-e.html 1

\bibitem{Gadzooks}
J.F. {Beacom}, M.R. {Vagins}, Phys. Rev. Lett. \textbf{93}(17), 171101 (2004)

\bibitem{DSNB} J. F. Beacom and L. E. Strigari, Phys. Rev. C 73, 035807 (2006), 
M. Wurm et. al., Phys. Rev. D 75, 023007 (2007) 

\bibitem{SN1987A-20th}
\verb3 http://sn1987a-20th.physics.uci.edu/ 3

\bibitem{Fogli} G.L. Fogli, E. Lisi, A. Mirizzi and D. Montanino, JCAP 0504, 002 (2005)

\bibitem{NNN06}
\verb7 http://neutrino.phys.washington.edu/nnn06/ 7

\bibitem{Learned_GEO} J. G. Learned, S. T. Dye and S. Pakvasa,
''Neutrino Geophysics Conference Introduction'', 	
{\it Earth, Moon, and Planets} \textbf{99} (2006) 1

\bibitem{HanoHano}
\verb7 http://www.phys.hawaii.edu/~sdye/hano.html 7


\bibitem{GigatonArray} J. G. Learned, ''White paper on Gigaton Array'',
\verb9 www.phys.hawaii.edu/~jgl/post/gigaton_array.pdf 9

\bibitem{Davis}
R. Davis, Jr. Phys. Rev. Lett. 12, 303 (1964) \\
J. N. Bahcall and R. Davis, Jr. Science 191, 264-267 (1976)

\bibitem{Gallex}
GALLEX-Collaboration: P. Anselmann et al.
Physics Letters B 357(1-2) (1995) 237-247\\
W. Hampel et al. Physics Letters B 388(2) (1996) 384-396\\
N. Bahcall, B. T. Cleveland, R. Davis et.al.
Phys. Rev. Lett. 40, 1351-1354 (1978)  
\bibitem{SNO}
The SNO Collaboration, Phys.Rev.Lett. 87 (2001) 071301
\bibitem{SK_sun}
S. Hirata et al., Phys. Rev. Lett. 65, 1297, 1301 (1990); 66, 9 (1991); 
Phys. Rev. D44, 2241 (1991).
\bibitem{SK_sn}
Hirata, K. S. et al. (Kamiokande), Phys. Rev. D38 (1988) 448-458; 
Phys. Rev. Lett. 58 (1987) 1490-1493.
\bibitem{IMB}
Bionta, R. M. et al. (IMB), Phys. Rev. Lett. 58 (1987) 1494.
\bibitem{LSD}
Galeotti, P. et al., Helv. Phys. Acta 60 (1987) 619-628.

\bibitem{Baksan}
Alekseev, E. N., Alekseeva, L. N., Volchenko, V. I., Krivosheina, I. V., JETP Lett. 45 (1987) 589-592.\\
Pisma Zh. Eksp. Teor. Fiz. 45, 461-464 (1987)\\
Chudakov, A. E., Elensky, Ya. S., Mikheev, S. P., JETP Lett. 46 (1987) 373-377.\\
Pisma Zh. Eksp. Teor. Fiz. 46, 297 (1987).\\
Alekseev, E. N., Alekseeva, L. N., Krivosheina, I. V., Volchenko, V. I., Phys. Lett. B205 (1988) 209-214. 

\bibitem{Bahcall}
J. N. Bahcall and M. H. Pinsonneault, Rev. Mod. Phys. 64, 885 (1992)\\
J. N. Bahcall and R. N. Ulrich, Rev. Mod. Phys. 60, 297 (1988)\\
S. Turck-Chieze and I. Lopes, Astrophys. J. 408, 347 (1993)

\bibitem{MPA}
H.-Th. Janka et. al. astro-ph/0612072

\bibitem{Burrows}
A. Burrows, Nature, 403, 727 (2000)

\bibitem{Mezzacappa}   
J. Blondin, A. Mezzacappa, Nature 445, 58-60 (4 January 2007)

\bibitem{Yamada}
K. Kotake, S. Yamada, and K. Sato 
Phys. Rev. D, 68, 044023, (2003)

\bibitem{Bethe} Bethe, H. A., Rev. Mod. Phys. 62 (1990) 801-866.

\bibitem{Pons} 
J. A. Pons, A. W. Steiner, M. Prakash, and J. M. Lattimer,
Phys. Rev. Lett. \textbf{86} (2001) 5223

\bibitem{RedGiants}
G. Raffelt \& A. Weiss, Astron. Astrophys. 264 (1992) 536-546. 

\bibitem{WDcool} L. G. Althaus, E. Garcia-Berro, 
J. Isern, A. H. Corsico, A\&A 441, 689-694 (2005) 

\bibitem{IaSmouldering}
 W. Hillebrandt and J. C. Niemeyer, Annual Review of Astronomy and Astrophysics
\textbf{38} (2000) 191-230 

\bibitem{HaenselRev} {{Yakovlev}, D.~G. and {Kaminker}, A.~D. and {Gnedin}, O.~Y. and 
        {Haensel}, P.}, Physics Reports \textbf{354}  1 (2001)

\bibitem{Heger_rev} S. E. Woosley, A. Heger, \& T. A. Weaver,  
RMP \textbf{74} (2002) 1015

\bibitem{14C} S. Schönert et al. (BOREXINO Collaboration), physics/0408032 
[Nucl. Instrum. Meth. A (to be published)].

\bibitem{Ratkovic}
S.~Ratkovic, S.I. Dutta, M.~Prakash, Phys. Rev. D \textbf{67}(12), 123002 (~21) (2003),

\bibitem{Dutta}
S.I. {Dutta}, S.~{Ratkovi{\'c}}, M.~{Prakash}, 
Phys. Rev. D \textbf{69}(2), 023005 (2004)

\bibitem{MOK}
M.~Misiaszek, A.~Odrzywolek, M.~Kutschera, 
Phys. Rev. D \textbf{74}(4), 043006 (2006)

\bibitem{PDBook}
W.M. {Yao}, C.~{Amsler}, D.~{Asner}, R.~{Barnett}, J.~{Beringer}, P.~{Burchat},
  C.~{Carone}, C.~{Caso}, O.~{Dahl}, G.~{D'Ambrosio} et~al., {Journal of
  Physics G} \textbf{33} (2006) 1, \verb1 http://pdg.lbl.gov 1

\bibitem{Itoh_VIII}
N.~{Itoh}, A.~{Nishikawa}, Y.~{Kohyama}, Astrophys. J \textbf{470}, 1015 (1996)

\bibitem{Esposito}
S.~{Esposito}, G.~{Mangano}, G.~{Miele}, I.~{Picardi}, O.~{Pisanti}, Nuclear
  Physics B \textbf{658}, 217 (2003)

\end{thebibliography}
\end{document}